# Colloidal Surfaces with Boundaries, Apex Boojums and Nested Elastic Self-Assembly of Nematic Colloids

Sungoh Park,[1] Qingkun Liu,[1] and Ivan I. Smalyukh[1,2,3,*]

[1]Department of Physics, University of Colorado, Boulder, CO 80309, USA
[2]Department of Electrical, Computer and Energy Engineering, Soft Materials Research Center and Materials Science and Engineering Program, University of Colorado, Boulder, CO 80309, USA
[3]Renewable and Sustainable Energy Institute, University of Colorado and National Renewable Energy Laboratory, Boulder, CO 80309, USA
[*]ivan.smalyukh@colorado.edu

**Self-assembly of colloidal particles is poised to become a powerful composite material fabrication technique, but remains challenged by a limited control over the ensuing structures. We develop a new breed of nematic colloids that are physical analogs of a mathematical surface with boundary, interacting with the molecular alignment field without inducing defects when flat. However, made from a thin nanofoil, they can be shaped to prompt formation of self-compensating defects that drive pre-programmed elastic interactions mediated by the nematic host. To show this, we wrap the nanofoil on all triangular side faces of a pyramid, except its square base. The ensuing pyramidal cones induce point defects with fractional hedgehog charges of opposite signs, spontaneously align with respect to the far-field director to form elastic dipoles and nested assemblies with tunable spacing. Nanofoils shaped into octahedrons interact as elastic quadrupoles. Our findings may drive realization of low-symmetry colloidal phases.**

Surface point defects, dubbed "boojums", arise in many condensed matter systems and often define their physical behavior [1-7]. For example, in nematic liquid crystal (LC) colloids and in confined LCs, topological characteristics of the particle-induced boojums are determined by the surface topology of colloidal inclusions [7,8]. Elastic interactions of particles in LCs, in turn, depend on these defects and their locations, ultimately leading to different forms of assembly [3-9]. By changing genus of colloidal particles, one can pre-determine the patterns, winding numbers, and hedgehog charges of induced surface defects [7,8]. However, surface genus is not the only means of controlling defects and self-assembly as even spherical particles can induce different topological singularities, including Saturn-ring disclination loops, hedgehogs, and boojums, giving rise to dipolar, quadrupolar, and hexadecapolar interactions [3-5,7-9]. Geometric shape of particles was also identified as a means of pre-determining elastic dipolar or quadrupolar interactions [5], albeit only for colloidal surfaces without boundaries and with conventional defects [10].

We introduce a new breed of LC colloids made from nanofoil, which are physical analogs of mathematical surfaces without boundary and induce no defects when flat. However, we uncover particle-induced boojums with fractional geometry-defined hedgehog charges of opposite signs when these surfaces without boundary are shaped into hollow pyramids without a base dubbed "colloidal pyramidal cones" (CPCs). The boojums localize near the inner and outer apex points of CPCs and induce elastic dipole moments $\mathbf{d}_e$ along the base-tip vectors $\mathbf{b} \| \mathbf{d}_e$. The CPCs spontaneously align with $\mathbf{b}$ either parallel or perpendicular to the far-field director $\mathbf{n}_0$ and form nested self-assemblies with tunable base-to-base distances. Nanofoils shaped into octahedrons also spontaneously align and induce apex boojums but exhibit quadrupolar elastic interactions. Despite the complex geometry and topology, colloidal interactions between the nanofoils display electrostatic analogy, potentially enabling elasticity-mediated self-assembly of mesoscopic composites pre-designed by three-dimensional shaping of colloidal surfaces without boundary.

To fabricate CPCs, flat discs and other particles made from nanofoil, we combine photolithography and thin film deposition (supplementary information and Figs. S1 and S2) [7,11]. Particles within the initial aqueous dispersions are functionalized by thiol-terminated methoxy-poly(ethylene glycol) (mPEG-SH, Jenkem Technology), re-dispersed in isopropanol and then added to a nematic 4-cyano-4'-pentylbiphenyl (5CB, Frinton Labs, Inc.), followed by solvent evaporation. The ensuing nematic dispersions are sandwiched between two glass plates treated to align the LC. By treating these plates with N,N-dimethyl-N-octadecyl-3-aminopropyl-trimethoxysilyl chloride (DMOAP) [8], we set perpendicular boundary conditions for the director **n**. By spin coating, baking and rubbing the polyimide PI-2555 (HD Microsystem) on the plates, we set tangential surface boundary conditions for **n**.

Polarizing, reflection and brightfield optical imaging, videomicroscopy and laser manipulations of particles were performed using a single integrated setup built around an inverted microscope IX81 (Olympus) [12]. To define initial conditions in probing particle interactions and to measure colloidal forces, we used optical tweezers based on a 1064 nm Ytterbium-doped fiber laser (YLR-10-1064, IPG Photonics) and a phase-only spatial light modulator (P512-1064, Boulder Nonlinear Systems) [12]. An Olympus 100× oil-immersion objective with a numerical aperture of 1.4 was used for both imaging and laser trapping. Motion of particles was recorded with a charge coupled device camera (Flea, PointGrey) at a rate of 15-60 fps and their lateral positions versus time were determined from captured image sequences using motion tracking plugins of ImageJ software (from NIH) with an accuracy of 7-10 nm [5,7,12].

Disc-shaped flat nanofoils spontaneously align with large-area faces parallel to **n**$_0$, while freely rotating around it, and interact with the surrounding LC depending on thickness $h_f$ of the gold foil (Fig. 1). They induce two surface boojum defects at the poles along **n**$_0$ when $h_f \approx 1\mu m$ (Fig. 1a), similar to the ones accompanying microspheres [3,4], but not when $h_f\approx100$nm (Fig. 1b). The thick-foil discs are homeomorphic to spheres with Euler characteristic $\chi=2$ and obey constraints of the Poincare-Hopf theorem by inducing surface defects (Fig. 1c), similar to colloidal microspheres [6-9, 13-15]. The surface boundary conditions for **n** at the edges of thin-foil discs are violated (Fig. 1d) and they interact with the surrounding **n(r)** as an orientable surface with a boundary, for which the Poincare-Hopf theorem cannot be applied. To minimize free energy, flat colloidal surfaces with boundaries and tangential anchoring spontaneously align along $\mathbf{n}_0$ without distorting it (Fig. 1b,d), unless rotated away from $\mathbf{n}_0$ by external forces. Overall, the LC-foil interactions are determined by a competition of bulk elastic and surface anchoring energies, which is characterized by the extrapolation length $\xi_e=K/W$, where K is the average elastic constant ($K\approx6.5$pN for 5CB) and W is the surface anchoring coefficient ($W\approx10^{-5}$J/m$^2$ for our PEG-coated gold foil surfaces; estimated as described for the gold surfaces functionalized with mPEG-SH in Ref. [16] and more generally in Ref. [17]). When $h_f<<\xi_e\approx650$nm, e.g. for $h_f\approx100$nm (Fig. 1b), the thin-foil particles behave as colloidal analogs of orientable surfaces with boundary that induce no defects (Fig. 1b,d). However, such colloidal surfaces can be morphed and folded into complex shapes, such as CPCs (Fig. 2), where pre-designed discontinuities in boundary conditions at vertices and edges distort **n(r)** and induce self-compensating defects that mediate elastic self-assembly, as we show below using nanofoil of $h_f\approx100$nm. Particles with $h_f\geq100$nm and micrometer-range overall dimensions (Fig. 2b) are mechanically rigid when dispersed in LCs, albeit nanofoil shapes can further morph when competing with LC elasticity at $h_f<100$nm and larger lateral dimensions, which, along with the colloidal behavior at lateral sizes approaching $\xi_e$, will be explored elsewhere. Below we use CPCs and octahedrons with 3.8μm bases and $h_f \approx 100$nm to exemplify the unexpected physical behavior of our colloidal analogs of mathematical surfaces with boundary.

CPCs spontaneously orient with their base-tip vectors $\mathbf{b} \perp \mathbf{n}_0$ or $\mathbf{b} \| \mathbf{n}_0$ and impose tangential anchoring on the surrounding $\mathbf{n}(\mathbf{r})$ (Fig. 2). The particle geometry causes director distortions visible in polarizing micrographs (Fig. 2c-f), with the boojum defects at the apex points of inner and outer pyramids of CPCs. To characterize $\mathbf{n}(\mathbf{r})$ around boojums at the interior and exterior tips of pyramids, we surround them by fragments of spheres $\sigma$ on the LC side and calculate an effective bulk charge $m_b = (1/4\pi) \int_\sigma d\theta d\phi \, \mathbf{n} \cdot [\frac{\partial \mathbf{n}}{\partial \theta} \times \frac{\partial \mathbf{n}}{\partial \phi}]$, where $\theta$ and $\phi$ are arbitrary coordinates on the $\sigma$-surface [7,18,19]. Despite the LC's nonpolar nature, we treat $\mathbf{n}(\mathbf{r})$ as a vector field within this procedure [8] (supplementary Fig. S3). Mapping vectorized $\mathbf{n}(\mathbf{r})$ onto a two-dimensional sphere $S^2$ does not fully cover it and the ratio of the covered and total areas of the unit sphere is the fractional charge $m_b$. Locally flat surfaces with tangential anchoring host boojums with $m_b = \pm 1/2$ that add to zero for surfaces without boundaries [6,7,19], $\Sigma_i m_{bi} = 0$. For CPC surfaces with boundaries, one also finds self-compensation of inner and outer boojums, which, for example, have $m_b = \pm(1-\cos\Omega)/2$ in the case of CPCs with $\mathbf{b} \perp \mathbf{n}_0$, where $\Omega$ is the effective cone half-angle characterizing colloidal surfaces near the apex points (Fig. 2c,d and supplementary Fig. S3). The signs of $m_b$ change to opposite with reversing the vectorized **n(r)** [8], but self-compensation persists. In the director field **n**$_s$(**r**) projected to the surface of the CPCs with **b**∥**n**$_0$, one finds two integer-strength two-dimensional defects at the interior and exterior pyramid apex points (Fig. 2e,f). Similar to the case of discs (Fig. 1), the CPCs become homeomorphic to spheres and induce defects that satisfy constraints of the Poincare-Hopf theorem when $h_f > \xi_e$, which will be explored in details elsewhere.

The $C_{4v}$ point symmetry group of CPCs holds in a nematic host for **b**∥**n**$_0$, but the particle-induced director distortions around CPCs reduce it to $C_{2v}$ for **b**⊥**n**$_0$ (Fig. 2 and supplementary Fig. S4). The ensuing elastic dipoles can be understood within the frameworks of electrostatic analogy and nematostatics [13-15]. To study them, we characterize their orientation-dependent anisotropic diffusion based on Brownian motion (supplementary Table S1) and then probe how they interact with microspheres with perpendicular anchoring (Fig. S5), for which the analogy with electrostatic dipoles is well-established [13,14]. CPCs with **b**∥**n**$_0$ exhibit dipolar interactions with the microspheres, consistent with the symmetry of distortions and elastic dipoles **d**$_e$∥**b** that they induce (Fig. S5). For example, CPCs and spheres separated along **n**$_0$ attract for parallel (Fig. S5a,b) and repel for anti-parallel **d**$_e$-vectors (Fig. S5c,d). CPCs with **d**$_e$∥**b**⊥**n**$_0$ resemble elastic dipoles formed by triangular and pentagonal prisms with such dipole moment orientations relative to **n**$_0$ [5,15]. Similar to the conventional nematic colloids, dipolar interactions between pairs of nanofoil-based colloidal analogs of mathematical surfaces with boundary shaped as CPCs emerge from the minimization of free energy due to elastic distortions induced by the individual particles, as we demonstrate below.

CPCs with $\mathbf{b}\perp\mathbf{n}_0$ always attract (Fig. 3), in both planar and homeotropic cells. When released from laser tweezers at different separation distances $r_{cc}$, azimuthal orientations of $\mathbf{d}_e$ around $\mathbf{n}_0$ and orientations of the center-to-center vector $\mathbf{r}_{cc}$ relative to $\mathbf{n}_0$ (Fig. 3a,b), CPCs undergo translational and rotational (around $\mathbf{n}_0$) motion to co-align their $\mathbf{d}_e$ moments, attract and self-assemble (Fig. 3c,d), regardless of the initial conditions. Time dependence matches that expected for dipolar interactions, $r_{cc}(t) = (r_0^5 - 5\alpha_d t)^{1/5}$ (Fig. 3a), where $r_0 = r_{cc}(0)$ is the initial distance and $\alpha_d$ is a fitting coefficient [5], albeit the rotations and nonlinear effects at small $r_{cc}$ can cause departures from this behavior. The ensuing nested self-assemblies of CPCs with co-aligned $\mathbf{d}_e\perp\mathbf{n}_0$ have average inter-base distances ≈0.5µm in homeotropic and ≈0.9µm in planar cells. This confinement-dependent separation between CPC bases is governed by near-field interactions associated with $\mathbf{n}(\mathbf{r})$-distortions at the apex boojums and bases of CPCs, which can be tuned further by varying cell thickness and applying fields. Pairs of CPCs with $\mathbf{d}_e\|\mathbf{n}_0$ and anti-parallel $\mathbf{d}_e$ repel at $\mathbf{r}_{cc}\|\mathbf{n}_0$ but attract at $\mathbf{r}_{cc}\perp\mathbf{n}_0$ while CPCs with co-aligned, parallel $\mathbf{d}_e$ exhibit an opposite behavior (Fig. 4a-c). Angular dependencies of elastic forces probed using a combination of optical tweezers and videomicroscopy, such as the ones shown in Fig. 4c, are consistent with their dipolar anisotropic nature. The ensuing nested self-assemblies of CPCs with co-aligned $\mathbf{d}_e\|\mathbf{n}_0$ have inter-base distances $\approx 1\mu m$ (Fig. 4d,e).

When LC is quenched from an isotropic state to a mesophase, the nucleation process at this first-order transition prompts an assembly of octahedrons formed by pairs of CPCs with oppositely aligned $\mathbf{b}$ and matched bases (Fig. 5). We find octahedrons both with $\mathbf{b}\|\mathbf{n}_0$ (Fig. 5a,b) and $\mathbf{b}\perp\mathbf{n}_0$ (Fig. 5c,d), albeit this alignment of constituent CPCs is inconsequential for the octahedron's elastic multipole nature. The symmetry of the octahedron colloidal particles is reduced from $O_h$ in an isotropic medium to $D_{4h}$ in the nematic LC with its uniaxial ground-state symmetry $D_{\infty h}$ due to the particle-distorted $\mathbf{n}(\mathbf{r})$, which is tangential to octahedron's surface and matches $\mathbf{n}_0$ at large distances (supplementary Fig. S4). The colloidal octahedrons repel both when $\mathbf{r}_{cc}\|\mathbf{n}_0$ and $\mathbf{r}_{cc}\perp\mathbf{n}_0$, but attract at intermediate angles (Fig. 5e,f). Both pair interactions and self-assemblies of colloidal octahedrons in the LC are consistent with their quadrupolar nature (Fig. 5e,f). The formation of hollow octahedrons is an example of colloidal assembly that leads to topological transformation of two CPC surfaces with boundary into a single closed surface without boundary, which then becomes compliant with the Poincare-Hopf theorem. Indeed, surfaces of octahedrons have no boundaries (Fig. 5) and the two boojums at vertices along $\mathbf{n}_0$ have their +1 winding numbers of the singularities in $\mathbf{n}_s(\mathbf{r})$ adding to octahedron's $\chi=2$ while the defects on other vertices and edges self-compensate in both interior and exterior of octahedrons (supplementary Fig. S3b). Similar to the case of CPCs, the fractional $m_b$ of the apex boojums self-compensate, $\Sigma_i\, m_{bi} = 0$.

By analyzing videomicroscopy-based $r_{cc}(t)$ data, such as the ones in Fig. 3a, we determine the relative particle velocity $v_{cc} = \partial r_{cc}/\partial t$ and then the elastic force from its balance with the viscous drag force while neglecting the inertia effects [3-5, 9]. As the particles interact, we find maximum elastic inter-CPC and inter-octahedra forces in the range 1-55pN and the potential energies $(0.5-10)\times10^3 k_B T$ that drive their self-assembly (supplementary Table S2), comparable to those of elastic multipoles due to other LC inclusions with similar dimensions [3-5] (here $k_B$ is the Boltzmann constant and T is the absolute temperature). By lowering the strength of these interactions to a few $k_B T$, say through reducing particle dimensions or by balancing elastic forces with electrostatic interactions, one can design and realize low-symmetry molecular-colloidal phases with varying degrees of fluidity and both orientational and positional order.

To conclude, we have developed nanofoil nematic colloidal analogs of a mathematical surface with boundary that induce no defects when flat but can be morphed to complex shapes that controllably generate self-compensating defects. CPCs and octahedrons made from the thin foil induce fractional boojums and exhibit nested and other symmetry-defined forms of robust elastic self-assembly. These nematic colloids may enable the realization of soft matter composites with low-symmetry structure, facile response to external stimuli [3, 5, 13,20], and properties, such as plasmonic, that arise not only from the superposition of properties due to individual constituent particles, but also from their precise placement with respect to each other within the self-assembled composite. Beyond elastic interactions, pinning of surface defects at the vertices may help defining "valence" in defect-mediated self-assemblies of particles [21,22]. Since extensions of the Poincare-Hopf theorem for surfaces with boundaries are known only for well defined conditions on the boundaries but not in general, experimental realization of such surfaces in LC-colloidal systems may allow for probing the interplay between the surface topology and that of induced defects in physical fields. Because all surfaces are characterized (up to homeomorphism) by their genus, orientability, and number of boundary components, as stated by the classification theorem [18], the colloidal system we have developed further expands the scope of experimental topology. Our study may also enable colloidal analogs of nonorientable surfaces, such as Möbius strips [23]. CPCs made from magnetic materials, such as nickel, may allow for estimating the free energy barrier associated with the particle rotation between the two stable orientations, which we will pursue elsewhere. Moreover, electric [24] and magnetic switching between these orientations may lead to bistable electro-optic devices.

We thank P. Chen, X. Hu, K. Janani, T. Lee, A. Mickelson, and B. Senyuk for discussions and technical assistance. This work was supported by the NSF grant DMR-1410735.

**Figures:**

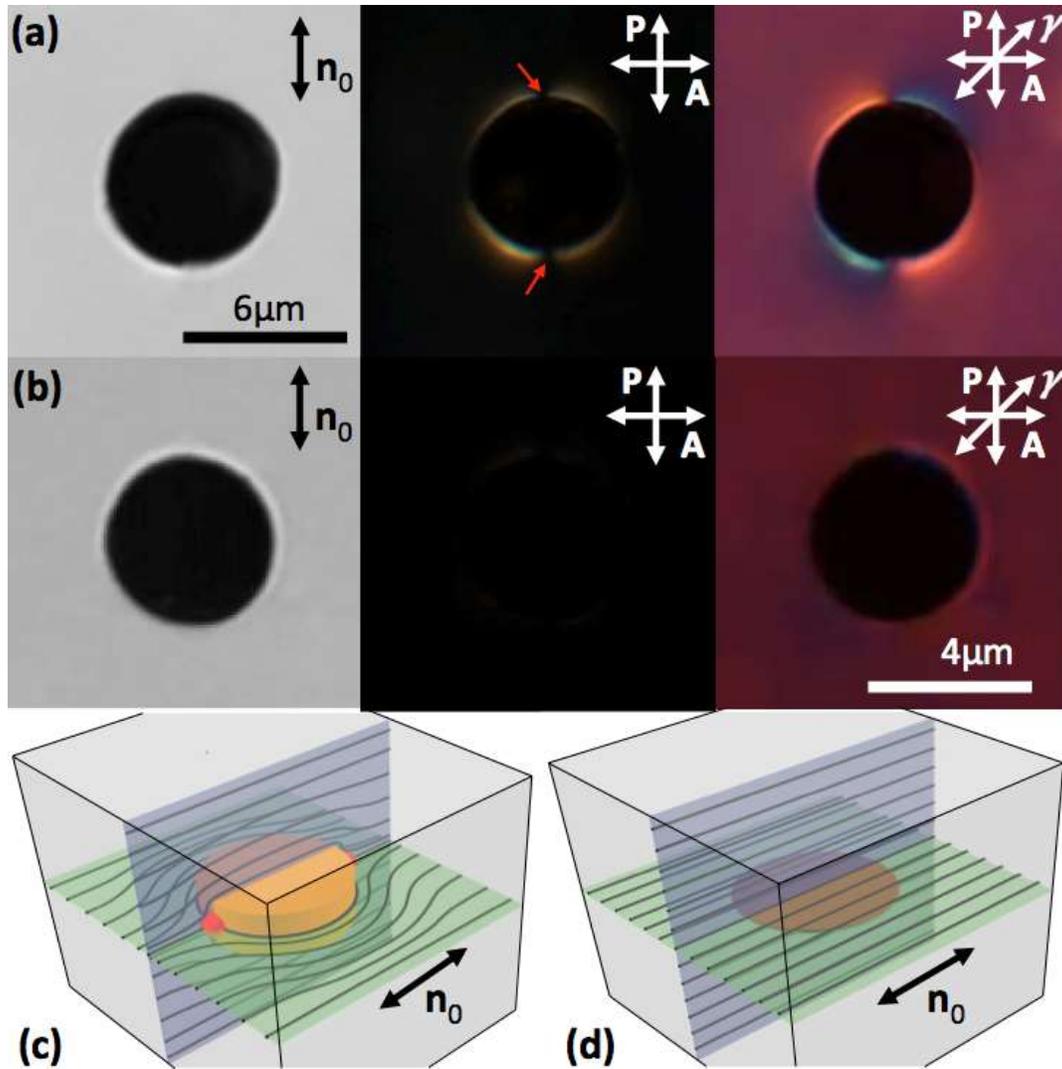

**Fig. 1.** Nematic colloidal analogs of mathematical surfaces (a,c) with and (b,d) without boundary. (a,b) Optical micrographs of discs made of (a) thick foil with $h_f \approx 1 \mu m$ and (b) thin foil with $h_f \approx 100 nm$ obtained in bright-field (left) and polarizing imaging modes without (middle) and with (right) a 530nm phase retardation plate with a fast axis $\gamma$ inserted between the crossed polarizer (P) and analyzer (A); red arrows indicate boojums. (c,d) Schematic illustrations of $\mathbf{n}(\mathbf{r})$ around the (c) thick and (d) thin discs, with the boojums in (c) depicted as red hemispheres.

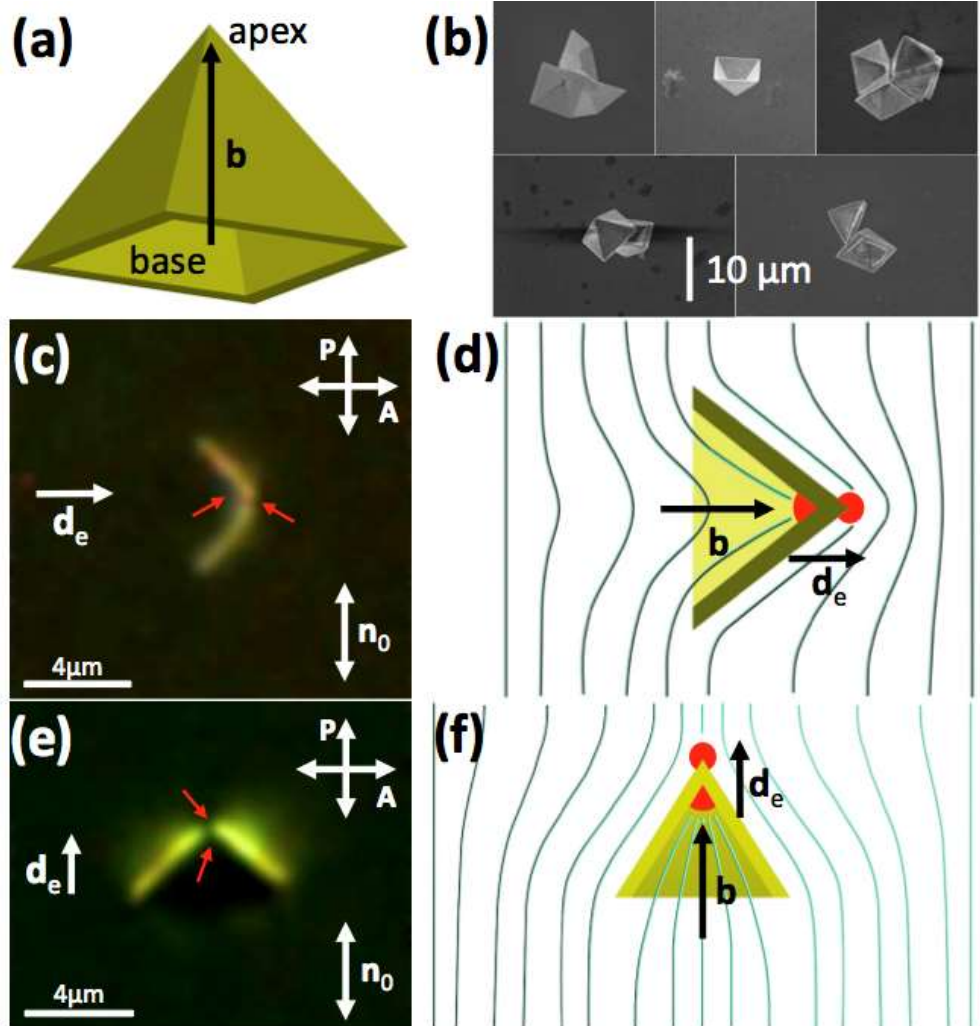

**Fig. 2.** CPCs in a nematic LC. (a) Schematic of a CPC. (b) Scanning electron microscopy of CPCs made from gold foil with (from top-left to bottom-right) $h_f \approx 100$nm, $h_f \approx 200$nm, $h_f \approx 300$nm, $h_f \approx 1\mu$m, $h_f \approx 1.1\mu$m. (c) Polarizing optical micrograph and (d) schematic of $\mathbf{n}(\mathbf{r})$ and defects around particles that orient with $\mathbf{b} \perp \mathbf{n}_0$. (e) Polarizing optical micrograph and (f) schematic of $\mathbf{n}(\mathbf{r})$ and defects around CPCs with $\mathbf{b} \| \mathbf{n}_0$. Complementary red fragments of spheres in (d,f) show fractional boojums with undefined $\mathbf{n}(\mathbf{r})$. Orientations of crossed polarizer (P) and analyzer (A) and $\mathbf{n}_0$ are shown using double arrows. In (c,e), $h_f \approx 100$nm and red arrows indicate boojums.

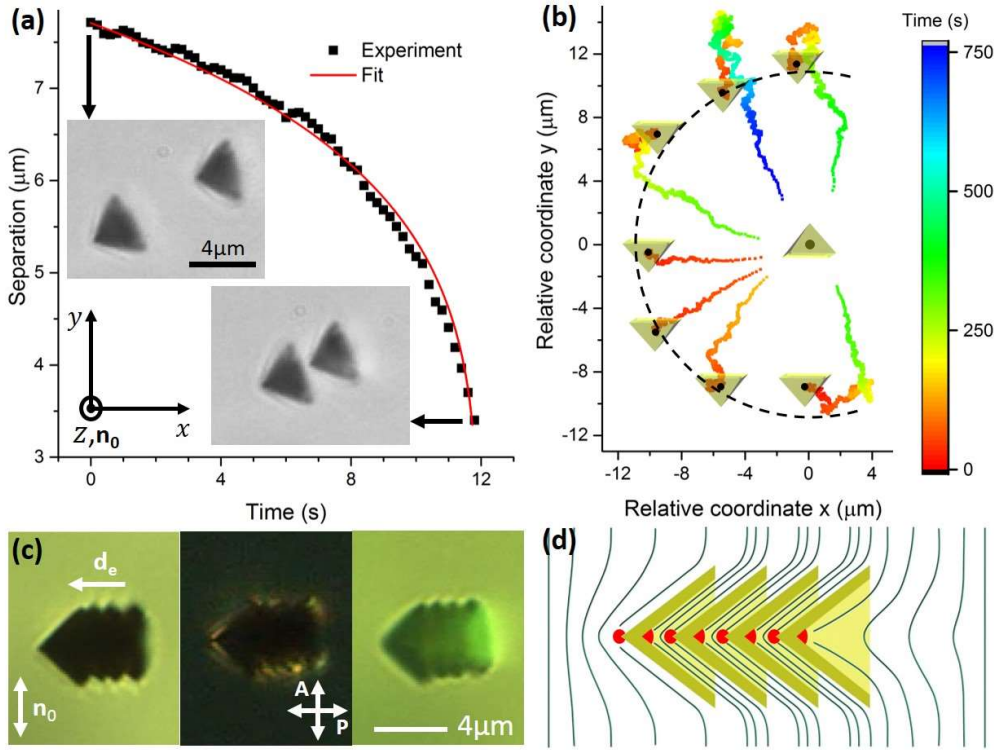

**Fig. 3.** Self-assembly of CPCs with $\mathbf{b}\perp\mathbf{n}_0$. (a) Experimental (squares) $r_{cc}(t)$ for CPCs with co-aligned $\mathbf{d}_e$ vectors and its fit (red line) shown for $r_0$=7.7μm and $\alpha_d$=456.9 μm$^5$/s. (b) Color-coded time trajectories of two CPCs with antiparallel $\mathbf{d}_e$ released simultaneously from optical traps in a homeotropic cell, demonstrating that particles rotate around the vertical $\mathbf{n}_0$ (orthogonal to the image plane and glass substrates) and translate to always attract and form a nested self-assembly despite being released with anti-parallel initial orientations of $\mathbf{d}_e$. The trajectories were constructed from videomicroscopy data, such as the ones shown in the supplementary video S1. Relative positions of the particles released from laser traps are shown with filled black circles atop of CPCs. (c) Brightfield (left), polarizing (middle), and reflection-mode (right) optical micrographs of a nested assembly of CPCs in a planar cell. The equilibrium inter-base distance within the nested assembly is ≈0.9 μm. (d) Schematic of $\mathbf{n}(\mathbf{r})$ around a nested colloidal assembly of CPCs. $h_f$≈100nm.

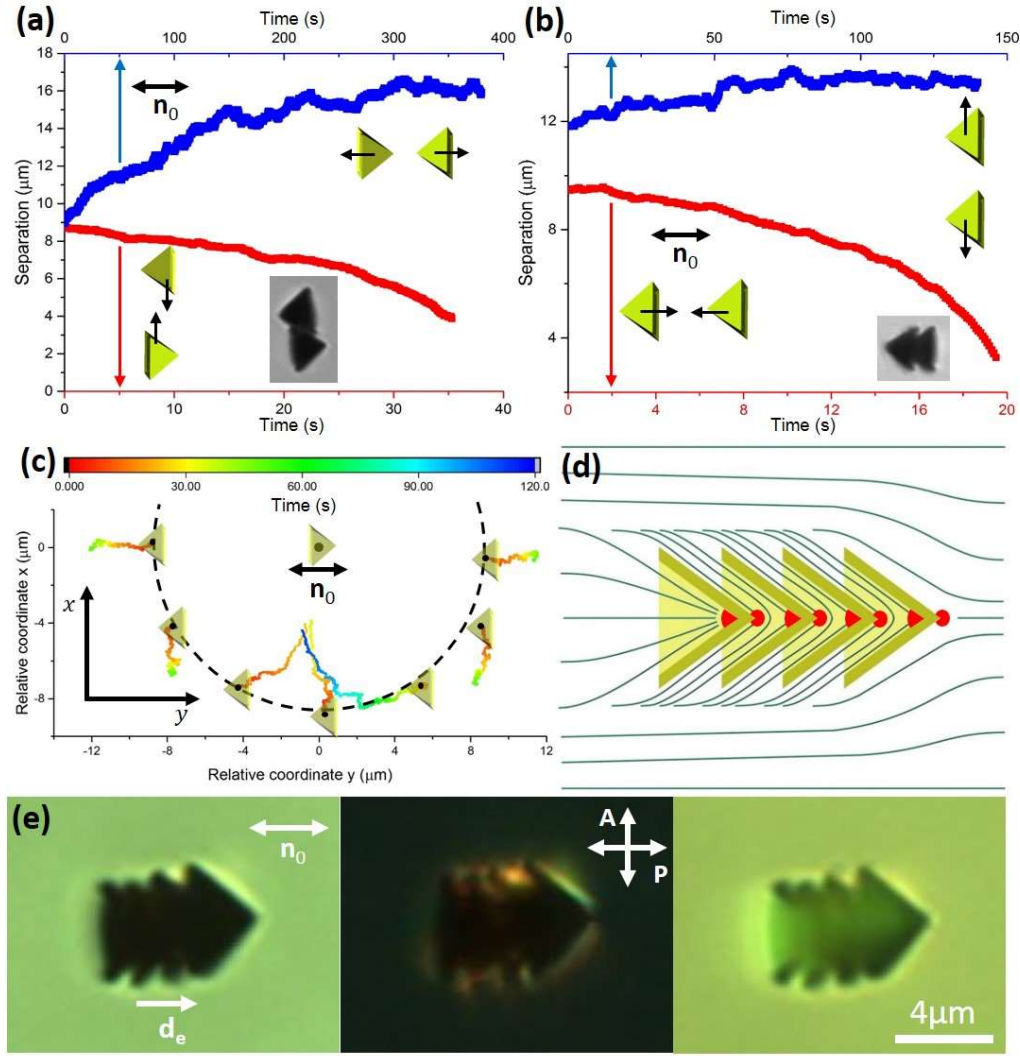

**Fig. 4.** Pair interactions and self-assembly of CPCs with **b**∥**n**$_0$. (a) Experimental $r_{cc}(t)$ for CPCs with antiparallel **b**, repelling at **r**$_{cc}$∥**n**$_0$ (blue curve) and attracting at **r**$_{cc}$⊥**n**$_0$ (red curve). (b) Experimental $r_{cc}(t)$ for CPCs with parallel **b**, attracting at **r**$_{cc}$∥**n**$_0$ (red curve) and repelling at **r**$_{cc}$⊥**n**$_0$ (blue curve). Black arrows in (a,b) indicate directions of elastic forces. (c) Color-coded (according to the color scheme shown at the top of the figure part) time trajectories of CPCs released from optical traps at different angles between **r**$_{cc}$ and **n**$_0$, as constructed from videomicroscopy data such as the ones shown in the supplementary video S2. (d) Schematic and (e) brightfield (left), polarizing (middle), and reflection (right) optical micrographs of a nested assembly of CPCs. The CPC assemblies with **d**$_e$∥**n**$_0$ have an average inter-base distance of ≈1.0μm. $h_f$≈100nm.

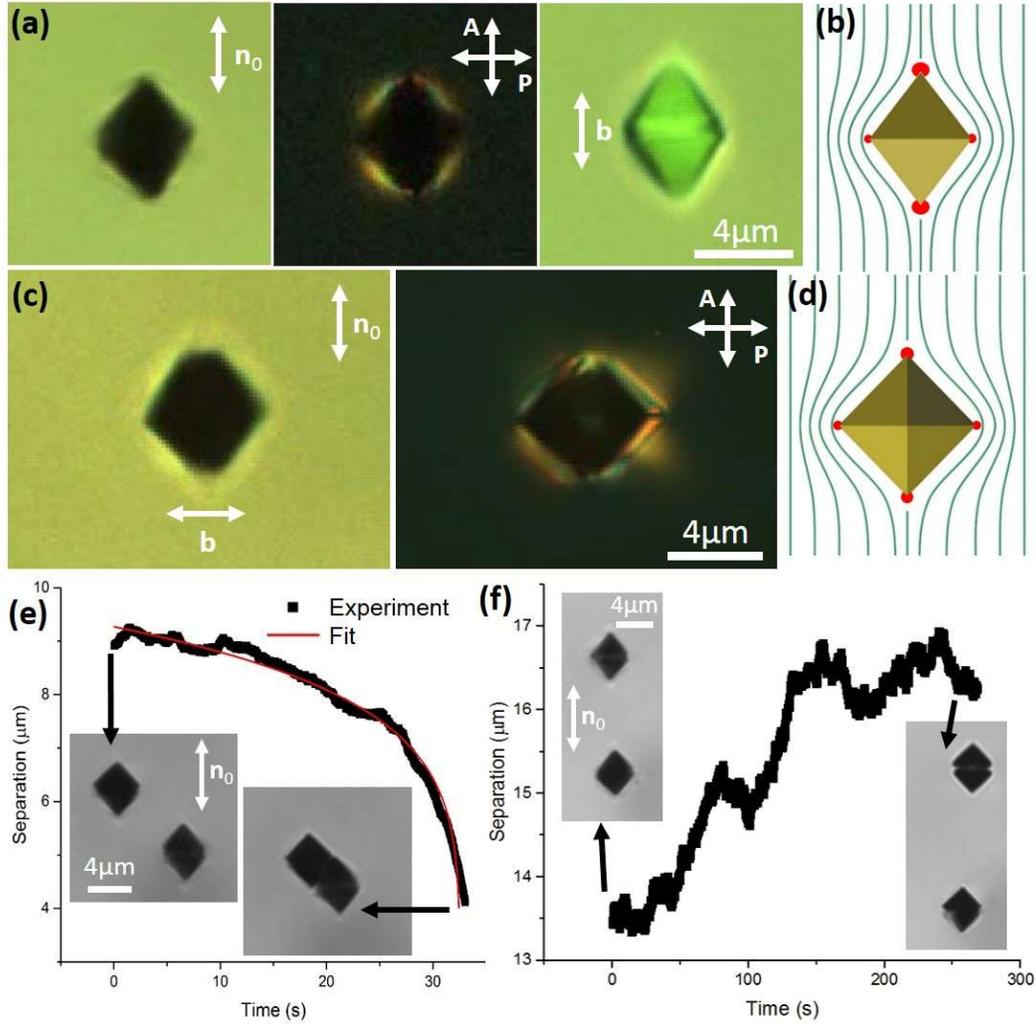

**Fig. 5.** Nematic colloidal octahedrons. (a) Brightfield (left), polarizing (middle), and reflection (right) micrographs showing an octahedron in a nematic LC, with the particle-induced distortion of **n(r)** and surface defects depicted in (b). (c) Brightfield (left) and polarizing (right) micrographs of an octahedron and (d) the corresponding **n(r)** and defects. (e,f) $r_{cc}(t)$ for two colloidal octahedrons with **r**$_{cc}$ at ≈45° to **n**$_0$ (e) and along it (f) obtained from the videomicroscopy data such as the ones shown in the supplementary video S3. The red curve is a fit with $r_{cc}(t) = (r_0^7 - 7\alpha_q t)^{1/7}$ expected for quadrupoles, yielding $r_0$=9.3µm and $\alpha_q$=2.6×10$^4$ µm$^7$/s. Insets in (e) and (f) are brightfield micrographs of the interacting octahedrons extracted as frames from the same videos at the elapsed times indicated by arrows. $h_f$≈100nm.